\documentclass[twocolumn]{aastex631}

\usepackage{CJKutf8, color}

\newcommand{\kanji}[1]{\begin{CJK}{UTF8}{ipxm}(#1)\end{CJK}}
\shorttitle{Constraints on EOS from RX J1804}
\shortauthors{Dohi et al.}
\graphicspath{{./}{./figures/}}
\begin{document}

\title{Constraints on the Neutron-Star Structure from the Clocked X-Ray Burster 1RXS J180408.9$-$342058}

\correspondingauthor{Akira Dohi}
\email{akira.dohi@riken.jp}

\author[0000-0001-8726-5762]{A.~Dohi \kanji{土肥明}}
\affiliation{Astrophysical Big Bang Laboratory (ABBL), Cluster for Pioneering Research, RIKEN, Wako, Saitama 351-0198, Japan}
\affiliation{Interdisciplinary Theoretical and Mathematical Sciences Program (iTHEMS), RIKEN, Wako, Saitama 351-0198, Japan}

\author[0000-0002-0207-9010]{W.~B.~Iwakiri \kanji{岩切渉}}
\affiliation{International Center for Hadron Astrophysics, Chiba University, Chiba 263-8522, Japan}

\author[0000-0002-0842-7856]{N.~Nishimura \kanji{西村信哉}}
\affiliation{Astrophysical Big Bang Laboratory (ABBL), Cluster for Pioneering Research, RIKEN, Wako, Saitama 351-0198, Japan}
\affiliation{RIKEN Nishina Center for Accelerator-Based Science, Wako, Saitama 351-0198, Japan}

\author[0000-0003-0943-3809]{T.~Noda \kanji{野田常雄}}
\affiliation{Department of Education and Creation Engineering, Kurume Institute of Technology, Kurume, Fukuoka 830-0052, Japan}

\author[0000-0002-7025-284X]{S.~Nagataki \kanji{長瀧重博}}
\affiliation{Astrophysical Big Bang Laboratory (ABBL), Cluster for Pioneering Research, RIKEN, Wako, Saitama 351-0198, Japan}
\affiliation{Interdisciplinary Theoretical and Mathematical Sciences Program (iTHEMS), RIKEN, Wako, Saitama 351-0198, Japan}
\affiliation{Astrophysical Big Bang Group (ABBG), Okinawa Institute of Science and Technology Graduate University (OIST),
Tancha, Onna-son, Kunigami-gun, Okinawa 904-0495, Japan}

\author{M.~Hashimoto \kanji{橋本正章}}
\affiliation{Department of Physics, Kyushu University, Fukuoka 819-0395, Japan}



\begin{abstract}

Type-I X-ray bursts are rapid-brightening transient phenomena on the surfaces of accreting neutron stars (NSs). Some X-ray bursts, called {\it clocked bursters}, exhibit regular behavior with similar light curve profiles in their burst sequences. The periodic nature of clocked bursters has the advantage of constraining X-ray binary parameters and physics inside the NS. In the present study, we compute numerical models, based on different equations of state and NS masses, which are compared with the observation of a recently identified clocked burster, 1RXS J180408.9$-$342058. We find that the relation between accretion rate and recurrence time is highly sensitive to the NS mass and radius. We determine, in particular, that 1RXS J180408.9$-$342058 appears to possess a mass less than $1.7M_{\odot}$ and favors a stiffer nuclear equation of state (with an NS radius $\gtrsim12.7{\rm km}$). Consequently, the observations of this new clocked burster may provide additional constraints for probing the structure of NSs.

\end{abstract}

% §１：イントロ：３−５（６）
% §２：RX J1804 のXバースト観測
%    - 解析の違い
%       --> 制限する
%    - <Δt>とC1stの導出まで
% §３：モデルからEOSの制限
%    - EOSの制限の基礎→「経験則」η
%    - モデル、ηの算出
%    - 比較M/M
% §４：まとめ



\section{Introduction} 
\label{sec:intro}

Type I X-ray bursts are rapidly evolving transient events observed from X-ray binaries, which are triggered by explosive thermonuclear burning on the accreting surface of a neutron star (NS). Observationally, 115 X-ray bursters have been identified \citep{2020ApJS..249...32G}, and most of them show an irregular pattern in X-ray light curves. Some exceptional cases show that the recurrence time of a series of X-ray bursts is quite regular in a few sources, which are called clocked bursters. The most representative clocked burster is GS 1826$-$24 \citep[see, e.g.,][and references therein]{2017PASA...34...19G}, which was first discovered in 1989 by the Ginga satellite \citep{1989ESASP.296....3T}. Due to the regular property of light curves (e.g., the almost constant recurrence time $\Delta t$, peak luminosity, and burst duration), clocked bursters are helpful in probing the various physical properties of low-mass X-ray binaries (LMXBs) and in modeling X-ray bursts, \citep[e.g.,][for GS~1826$-$24]{2007ApJ...671L.141H, 2018ApJ...860..147M, 2020MNRAS.494.4576J}. The observational light curves of GS~1826$-$24 also have been used for constraining relevant nuclear physics properties such as nuclear reaction rates involved in unstable proton-rich nuclei \citep{2019ApJ...872...84M, 2021PhRvL.127q2701H, 2022ApJ...929...72L,2022ApJ...929...73L} and the equation of state (EOS) and cooling of the central NS \citep{2021ApJ...923...64D, 2022ApJ...937..124D}.

Recent observations indicate that another X-ray burster, 1RXS~J180408.9$-$342058 (hereafter, RX J1804), would show the property of the clocked burster. RX J1804 is an LMXB system found by ROSAT satellite in 1990 \citep{1999A&A...349..389V}, of which the first X-ray burst event was detected by INTEGRAL in 2012 \citep{2012ATel.4050....1C}. In \cite{2017MNRAS.472..559W} and \cite{2019ApJ...887...30F}, two epochs of X-ray bursts are observationally identified in hard and transitional X-ray states. As a distinctive feature, the observed recurrence time for each epoch is almost constant in each series of bursts, which may imply that RX J1804 may be a clocked burster.

%The shape of these light curves of RX J1804 in each epoch appear to be constant, which may be a clocked burster.

In this work, we investigate the physical properties of neutron stars, e.g., EoS, by modeling a newly observed clocked burster RX J1804. We use a general relativistic stellar evolution code \citep[the \texttt{HERES} code, described in][]{2023ApJ...950..110Z} covering the whole NS regions. Such a numerical code enables us to investigate the NS physics, e.g., the EOS and neutrino cooling processes. For NS EOSs, there are many experimental and observational constraints \citep[e.g.,][]{2022PTEP.2022d1D01S}, but still have large uncertainties. Our approach can constrain the NS EOS from astronomical observations through X-ray bursts. %because the first-principle calculation based on quantum chromodynamics needed for the exact description of the high-density matter in NSs is impossible due to the notorious sign problem (for a review, see \cite{2022PrPNP.12703991N}).

The present paper is organized as follows. In Section~\ref{sec:methods}, we summarize the observational properties of RX J1804 and the methods for our X-ray burst models. In Section~\ref{sec:results}, we present the results, compared with RX J1804 burst observations, and show the EOS and NS mass constraint. Section~\ref{sec:con} is devoted to conclusions.

%%Fiocchi+19's Analysis
%\subsection{Analysis by \cite{2019ApJ...887...30F}}

\section{Methods}
\label{sec:methods}
\subsection{Observations of clocked bursters RX J1804}
\label{subsec:eta}
%\ADcomment{On this Section, careful check based on Fiocchi et al. 2019 would be needed ($\rightarrow$ Iwakiri-sann)}

%%Procedure of model constraint
We present the method to constrain burst models from RX J1804 burst observations. First, we make numerical burst models with various $\dot{M}_{-9}$, in particular $\dot{M}_{-9}$--$\Delta t$ relations, where $\dot{M}_{-9}$ is the normalised accretion rate in units of $10^{-9}~M_{\odot}~{\rm yr}^{-1}$. Then, we can obtain $\dot{M}_{-9,{\rm 1st}}$ and $\dot{M}_{-9,{\rm 2nd}}$ from each $\Delta t$, respectively. For the observational $\Delta t$ and $\dot{M}_{-9}$ of RX J1804, we take the results of \cite{2019ApJ...887...30F}, which analyzed the quasi-simultaneous INTEGRAL, SWIFT, and NuSTAR observational data over a very broad energy band of 0.8--200 keV. Namely, assuming the presence of the Clocked X-ray bursters, the averaged $\Delta t$ for each epoch is calculated as\footnote{We average the values of peak time ($T_{\rm peak}$) for each burst, which are listed in Table 3 of \cite{2019ApJ...887...30F}.}
%Table 3 is fully considered and rigidly
%dt1 = 1.07095679+-0.02654971,
%dt2 = 2.20358205+-0.03614175,
%where we neglect the data errors of each T_peak (+-1) because we take values up to O(10s).
\begin{eqnarray}
\Delta t_{\rm 1st} &=& 2.20\pm0.04~\text{h}~,\label{eq:1}\\
\Delta t_{\rm 2nd} &=& 1.07\pm0.03~\text{h}~,\label{eq:2}
\end{eqnarray}
which determine $\dot{M}_{-9,{\rm 1st}}$ and $\dot{M}_{-9,{\rm 2nd}}$ for various burst models, respectively. 

The other observation of RX J1804 is the \textit{unabsorbed} persistent flux, which is given as follows \citep[from Table 2 in][]{2019ApJ...887...30F}:
\begin{eqnarray}
f_{\rm per,1st} &=& 
\left(45\pm13\right)\times10^{-10}~{\rm erg~cm^{-2}~s^{-1}}~,\label{eq:3}\\
f_{\rm per,2nd} &=& \left(55\pm13\right)\times10^{-10}~{\rm erg~cm^{-2}~s^{-1}} \label{eq:4}
\end{eqnarray}
for 1st and 2nd epochs, respectively. Since the persistent flux is directly proportional to the accretion rate, we can deduce the observational ratio of accretion rates between different epochs as~\footnote{Note that the best-fit column density obtained from spectral fitting is the same between two phases~\citep{2019ApJ...887...30F}, the ratio of accretion rate ratios can be simply obtained as the ratio between Eqs.~(\ref{eq:3}) and ~(\ref{eq:4}).}
\begin{eqnarray}
\frac{\dot{M}_{-9,{\rm 2nd}}}{\dot{M}_{-9,{\rm 1st}}}=1.2\pm0.5~\label{eq:5}.
\end{eqnarray}
Thus, we can judge the consistency of the model from $\dot{M}_{-9,{\rm 2nd}}/\dot{M}_{-9,{\rm 1st}}$ values. 
%In particular, we focus on the constraints on EOSs from 

A useful parameter to indicate the $\dot{M}_{-9}$--$\Delta t$ relations is the $\eta$ parameter, which is the power-law gradient being typically $\sim1$, as ~\citep[e.g.,][]{2016ApJ...819...46L}
\begin{eqnarray}
\Delta t \propto f_{\rm per}^{-\eta}=k\dot{M}^{-\eta}_{-9}~,
\label{eq:6}
\end{eqnarray}
where $k$ is a constant. In fact, many burst observations show that the $\dot{M}_{-9}$--$\Delta t$ relation matches with the power-law relationship with high accuracy. If $\eta=1$, there exists the critical mass of fuel for ignition $M_{\rm crit}\equiv\dot{M}^{\eta}_{-9}\Delta t$, but actually $\eta\neq1$ from most observations such as GS 1826$-$24 \citep[$\eta=1.05\pm0.02$][]{2004ApJ...601..466G} and MXB 1730$-$335~\citep[$\eta=0.95\pm0.03$,][]{2013MNRAS.431.1947B}. These facts imply that the amount of fuel for ignition varies with the accretion.

$\eta$ dependence of model parameter has been investigated in detail by \cite{2016ApJ...819...46L} with the 1D implicit hydrodynamics code with large reaction network$\sim$ 1300 nuclei, \texttt{KEPLER}~\citep{2004ApJS..151...75W}. Assuming the mass and radius of $M_{\rm NS}=1.4~M_{\odot}$ and $R_{\rm NS}=11.2~{\rm km}$, respectively, they calculated burst models with various $Z_{\rm CNO}$. They finally concluded that $\eta$ varies from 1.1 to 1.24 but is weakly sensitive to $Z_{\rm CNO}$. However, their burst models seem to be incompatible with the recent burst observations in RX J1804; from the empirical relation of Eq. (\ref{eq:6}), one can get
\begin{eqnarray}
\frac{\dot{M}_{-9,{\rm 2nd}}}{\dot{M}_{-9,{\rm 1st}}} = \left(\frac{\Delta t_{\rm 1st}}{\Delta t_{\rm 2nd}}\right)^{1/\eta} \simeq \left(2.05\right)^{1/\eta},
\label{eq:7}
\end{eqnarray}
which results in $\eta\gtrsim1.35$ for the case of \cite{2019ApJ...887...30F}. That is why we need to explore model parameters beyond \cite{2016ApJ...819...46L} in order to explain RX J1804 observations. As the candidates, we pay attention to the NS mass and EOS, whose uncertainties highly affect $\Delta t$ due to the simultaneous change of surface gravitational and neutrino cooling effects~\citep{2021ApJ...923...64D}.

%%\eta Parameter

%Narayan & Heyl (2003), which studies radius and cooling influenece on dMdt-dt relation (but not \eta dependence)
% \cite{2003ApJ...599..419N}

\subsection{Multi-zone X-ray-burst models}
\label{subsec:model}

To calculate X-ray burst models, we employ a multi-zone general-relativistic stellar evolution code, the \texttt{HERES} code, originally developed by \cite{1984ApJ...278..813F} and recent updates and comparison to \texttt{MESA} code are shown in \cite{2023ApJ...950..110Z}. We follow the quasi-hydrostatic thermal evolution of bursting NSs through successive bursts with nuclear burning of approximated (88 nuclei) reaction network for mixed hydrogen and helium burning \citep{2020PTEP.2020c3E02D}. The \texttt{HERES} code consistently includes the central NS with X-ray burst region, which allow us to investigate the dependency of EOS on X-ray burst light-curves \citep[see, e.g.,][]{2021ApJ...923...64D, 2022ApJ...937..124D}.

%In one-zone burst models, where nuclear burning shells are regarded as geometrically thin and the column density can be approximately expressed to be the ignition pressure over the surface gravity of NSs~\citep{1981ApJ...247..267F}, this small network can reproduce the large network with 897 nuclei in regards to the energy generation rate and ashes of hydrogen and helium within 40\% errors for $X/Y\gtrsim1$, where $X$ and $Y$ denote the mass fraction of hydrogen and helium, respectively~(\cite{2017KUPhD1806813}). 

For the data of reaction rates, we mostly adopt the JINA Reaclib database\footnote{https://reaclib.jinaweb.org} ver 2.0 \citep{2010ApJS..189..240C} except for ${}^{64}{\rm Ge}({\rm p},\gamma){}^{65}{\rm As}$ and ${}^{65}{\rm As}({\rm p},\gamma){}^{66}{\rm Se}$ rates \citep{2016ApJ...818...78L}, which have significant impacts on light curves.
%Recent studies have shown the updated reaction rates, such as ${}^{22}{\rm Mg}(\alpha,p){}^{25}{\rm Al}$ \citep{2021PhRvL.127q2701H}, ${}^{57}{\rm Cu}(p,\gamma){}^{58}{\rm Zn}$ \citep{2022ApJ...929...73L}, and ${}^{65}{\rm As}(p,\gamma){}^{66}{\rm Se}$~\citep{2022ApJ...929...72L}, may have impacts on light curves.
Since we mostly focus on the impacts of EOS on the recurrence time $\Delta t$ as a representative burst output parameter, we adopt the above reaction set and do not change in the present study. Details of the numerical procedure of systematic X-ray burst calculations are the same as \cite{2021ApJ...923...64D}. We describe the input parameters of our burst models.

As the NS microphysics, we employ three EOSs, i.e., Togashi \citep{2017NuPhA.961...78T}, LS220 \citep{1991NuPhA.535..331L}, and TM1e \citep{2020ApJ...891..148S}, among which the radius is significantly different \citep[see Figure~1 in][]{2021ApJ...923...64D}; there still remain uncertainties in the NS radius, which is $R_{\rm NS}\sim11$--$14$~{\rm km} with $1.4~M_{\odot}$.
%Many recent NS observations and experiments enable us to constrain the NS EOS, but there still remain some uncertainties of $R_{\rm NS}\sim11$--$14$~{\rm km} with $1.4~M_{\odot}$ stars, which our adopted EOSs satisfies.
The treatment of heating and cooling processes inside NSs is the same as \cite{2021ApJ...923...64D}, in which the conventional crustal heating process and the slow $\nu$ cooling process, mainly composed of the modified Urca process and Bremsstrahlung\footnote{The enhanced cooling by nucleon superfluidity can be ignorable, since the core temperature is much lower than the transition temperature \citep{2022ApJ...937..124D}.}, are implemented. The fast cooling process, such as the direct Urca process, could occur and have impacts on X-ray bursts in heavy NSs (and EOSs with the large symmetry energy) \citep{2022ApJ...937..124D}, but for simplicity we ignore this effect in the present study.

%We note that the uncertainties of crustal heating processes deduced from nuclear non-equilibrium reactions are not so important for EOS constraints from burst observations though they highly affect $\Delta t$. For the $\nu$ cooling

The observed light curves of RX J1804 show that these X-ray bursts are triggered by mixed H/He burning \citep{2017MNRAS.472..559W, 2019ApJ...887...30F}, implying $\dot{M}_{-9}\gtrsim1$ due to stability conditions of nuclear burning \citep{1981ApJ...247..267F, 1998ASIC..515..419B}. We accordingly choose $\dot{M}_{-9}=2-9$. Another crucial input parameter is the composition of accreted matter in particular for the metalicity $Z_{\rm CNO}$. As the RX J1804 locates in the global cluster in the Galactic bulge \citep{1999A&A...349..389V}, which implies relatively higher metalicity, we choose $Z_{\rm CNO} = 0.01, 0.015$ and $0.02$ including ${}^{14}{\rm O}$ and ${}^{15}{\rm O}$ at a ratio of 7 to 13. For the mass fraction of light elements, we fix to the solar abundance ratio, $X/Y=2.9$.

\section{Results}\label{sec:results}

\begin{figure*}[tp]
    \centering
    \includegraphics[width=\linewidth]{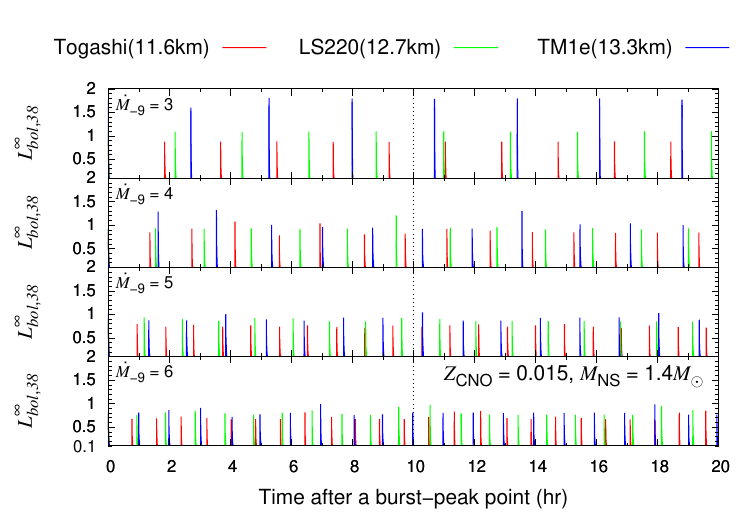}
    \caption{EOS dependence of averaged light curves during 20 hrs with $1.4 M_{\odot}$ NSs and $Z_{\rm CNO}=0.015$. The horizontal axis indicates the time where a time at the peak point is set to be zero, and the vertical axis indicates the bolometric luminosity in units of $10^{38}~{\rm erg~s}^{-1}$. $\dot{M}_{-9}=3$ (top), $\dot{M}_{-9}=4$ (second top), $\dot{M}_{-9}=5$ (second bottom), and $\dot{M}_{-9}=6$ (bottom).}
    \label{fig:lc}
\end{figure*}

Figure~\ref{fig:lc} shows X-ray burst light curves with various mass accretion rates and different EOSs. As shown by \cite{2021ApJ...923...64D}, a large-radius (stiff) EOS has a high $\Delta t$ and high peak luminosity for all mass accretion rates. This is because of the lower surface gravity, which requires more amount of fuel for the ignition. Although the $\Delta t$ significantly depends on the mass accretion rates, where higher $\dot{M}_{-9}$ shows a shorter $\Delta t$, we clearly find the above EOS dependence.

\begin{figure}[t]
    \centering
    \includegraphics[width=\linewidth]{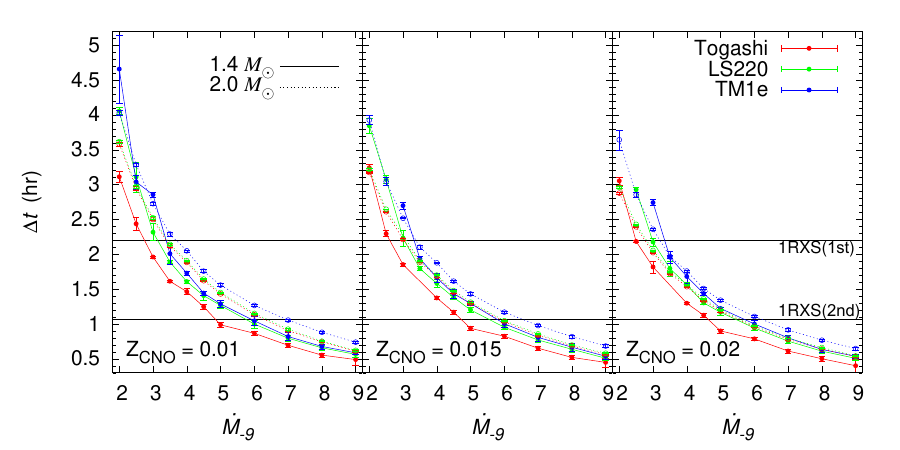}
    \caption{EOS dependence of averaged $\Delta t$ with 1 $\sigma$ errors. Solid curves indicate with 1.4 $M_{\odot}$ NSs while dotted curves 2 $M_{\odot}$ NSs. Metalicity $Z_{\rm CNO}$ is chosen to be 0.01 (left), 0.015 (middle), and 0.02 (right). The observational $\Delta t$ of RX J1804 for the 1st and 2nd epochs are plotted.}
    \label{fig:dt}
\end{figure}

The averaged $\Delta t$ values with different EOSs and $\dot{M}_{-9}$ are presented in Figure~\ref{fig:dt}. We can see that a stiffer EOS generally results in a higher $\Delta t$, except for the case of the lowest $\dot{M}_{-9} = 2$. Interestingly, in these low-$\dot{M}$ models, the dependence on the stiffness of the EOS becomes reversed. This phenomenon occurs due to the proximity of the peak luminosity to the Eddington luminosity at low $\dot{M}$. Consequently, the compressional heating luminosity caused by the gravitational release of NSs becomes more influential, and the effect of surface gravity becomes relatively more significant \citep[see also][]{2023ApJ...950..110Z}.

\begin{figure}[t]
 \centering
    \includegraphics[width=\linewidth]{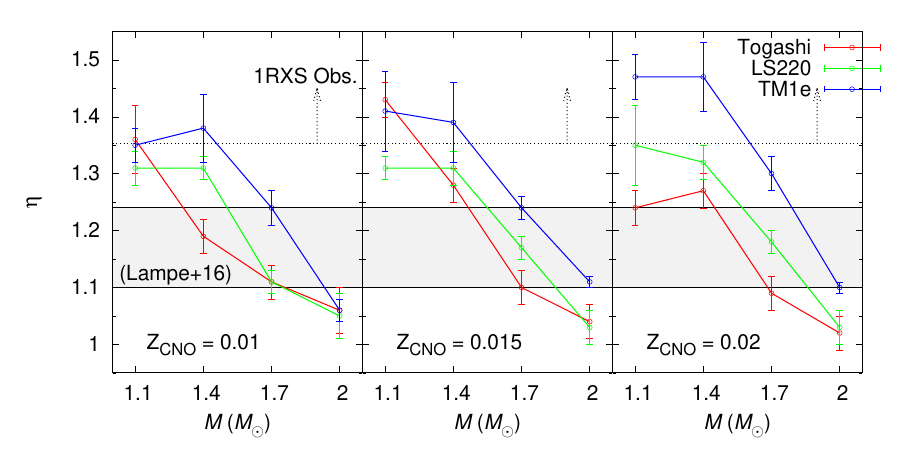}
    \caption{$\eta$ values as a function of NS masses. We also show the range obtained from previous work~\citep{2016ApJ...819...46L} and the lower limit of RX J1804 implied from Eq.~(\ref{eq:7}).}
    \label{fig:eta}
\end{figure}

By applying the fitting formula given by Eq.~(\ref{eq:6}) to the $\Delta t$--$\dot{M}_{-9}$ relations, as shown in Figure \ref{fig:dt}, we obtain the corresponding $\eta$ values shown in Figure \ref{fig:eta}. We note that the $\eta$ and $k$ values lie within less than 10\% error for all parameter regions of $Z_{\rm CNO}$, EOS, and NS mass. We find that stiffer EOSs tend to have higher $\eta$ values due to their lower $\Delta t$ values. Additionally, lower-mass models exhibit higher $\eta$ values. Remarkably, $\eta$ strongly reflects the stiffness of the EOS, unlike the $\Delta t$ value, which is also influenced by $\nu$ cooling effects. Therefore, $\eta$ could be a powerful indicator for constraining the NS structure.

In the previous works~\citep{2016ApJ...819...46L}, the $\eta$ values lie in the range of 1.1--1.24 assuming $M_{\rm NS}=1.4~M_{\odot}$ and $R_{\rm NS}=11.2~{\rm km}$, but we find that the original finding range can be extended if the mass and radius are changed. The closest models to \cite{2016ApJ...819...46L} for model parameters are those with $M_{\rm NS}=1.4~M_{\odot}$ and the Togashi EOS ($R_{\rm NS}=11.6~{\rm km}$). In such NSs, $\eta$ values with $Z_{\rm CNO}=0.015$ and $0.02$ are slightly higher than the range expected previously, though those with $Z_{\rm CNO}=0.01$ are consistent. However, since the influence of EOS uncertainties on $\eta$ is large as in Figure \ref{fig:eta}, the difference in radius between 11.2 km and 11.6 km may not be small. Moreover, we can confirm that the metalicity dependence on $\eta$ is not so large as found in \cite{2016ApJ...819...46L}.  Thus, these facts imply that our burst models are qualitatively consistent with \cite{2016ApJ...819...46L}.

\begin{figure}[t]
    \centering
    \includegraphics[width=0.8\linewidth]{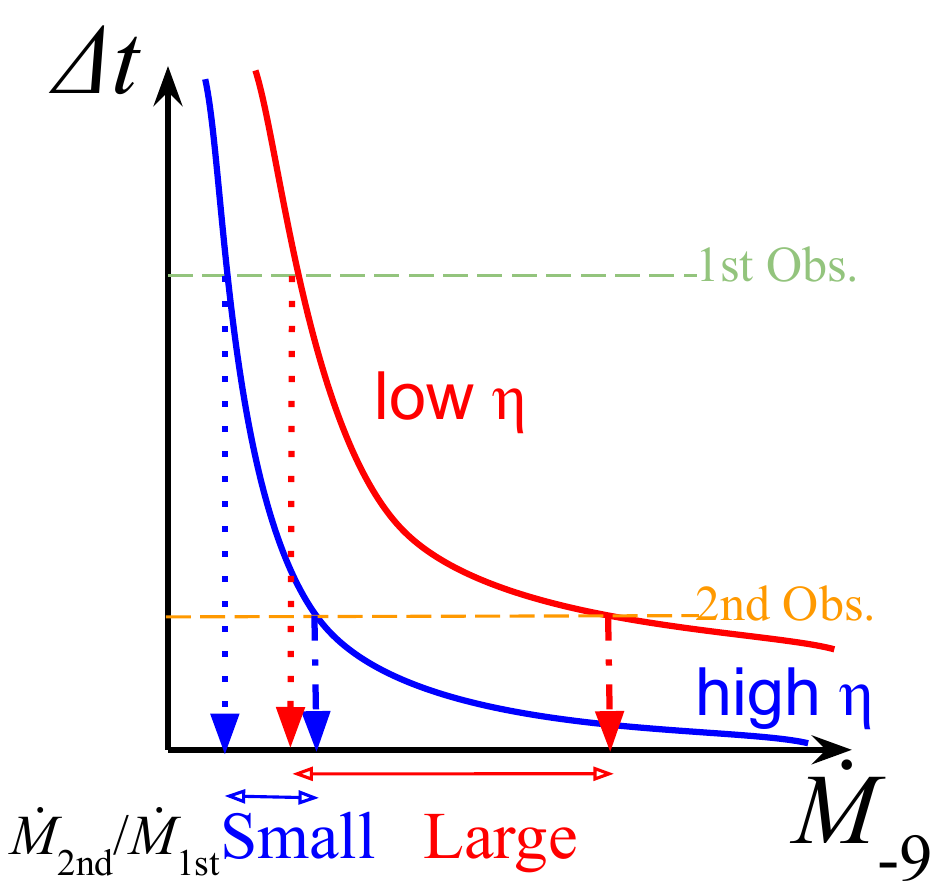}
    \caption{The schematic relation between $\eta$ and $\dot{M}_{-9,{\rm 2nd}}/\dot{M}_{-9,{\rm 1st}}$ with observed $\Delta t_{\rm 1st}$ and $\Delta t_{\rm 2nd}$ for RX J1804. 
    %(if the recurrence time for two epoch are determined, the accretion rate ratio is also determined through $\eta$ and burst models could be constrained from observations of each persistent flux.)
    }
    \label{fig:scheme}
\end{figure}

Except for $\Delta t$, the other observational factor for RX J1804 is the ratio of persistent flux between both epochs, i.e., accretion-rate ratio. To clarify the connection between $\eta$ and the accretion-rate ratio, we show the schematic picture in Figure \ref{fig:scheme}. First, we can find the crosspoints ($\dot{M}_{{-9},{\rm 1st}}$ and $\dot{M}_{{-9},{\rm 2nd}}$) between theoretical curves and RX J1804 observations for the 1st and 2nd epoch, respectively. Then, since $\eta>1$ and $\dot{M}_{-9}\gtrsim1$ for RX J1804, $\eta$ should be higher if the difference between crosspoints for the 1st and 2nd epoch, i.e., $\dot{M}_{{-9},{\rm 2nd}}/\dot{M}_{{-9},{\rm 1st}}-1$, becomes smaller. Thus, the accretion rate should have a negative correlation with $\eta$. 

\begin{figure}[t]
 \centering
    \includegraphics[width=\linewidth]{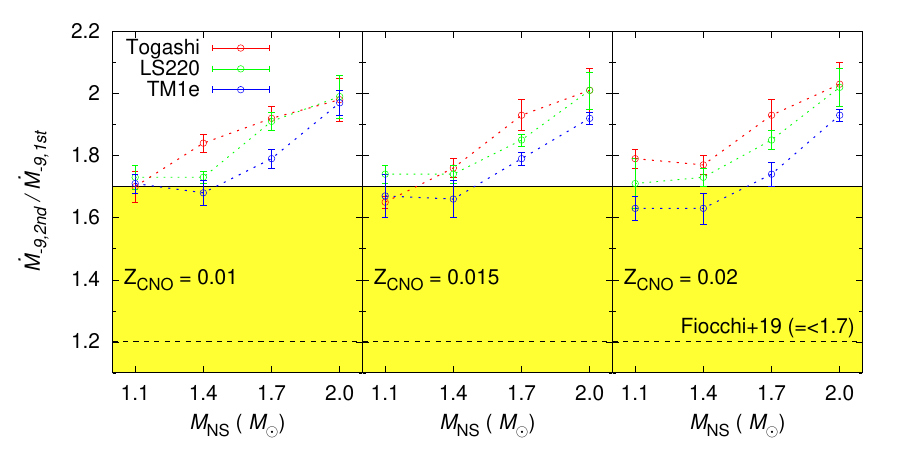}
    \caption{Same as Figure \ref{fig:dt}, but for the accretion-rate ratio between 1st and 2nd epoch. The observational constraint for RX J1804, Eq.~(\ref{eq:5}), is also plotted.}
    \label{fig:comp}
\end{figure}

Thus, we finally obtain the accretion-rate ratio as shown in Figure \ref{fig:comp}\footnote{Our calculation fixes the global accretion rate, which means that the local accretion rate per unit area $\frac{\dot{M}}{4 \pi R_{\rm NS}^2}$ varies with the EOS and mass. However, since the accretion-rate ratio is the same both with the global and local accretion rate, our conclusion holds.}. If the NS is more compact, the accretion-rate ratio tends to be higher, as we explain above. By comparing the observational constraint for RX J1804, i.e., Eq.~(\ref{eq:5}), one can see that RX J1804 should be light with at least $M_{\rm NS}<1.7~M_{\odot}$. This result may be peculiar in LMXBs because typical accreting NSs always get the mass from the companion in a long timescale ($\sim$ Gyr), and finally tend to become heavy. In fact, most observations show that accreting NSs in LMXBs are heavy~\citep{2016ARA&A..54..401O,2018MNRAS.478.1377A} \citep[see also][]{2022ApJ...934L..17R}. Our results thus imply that RX J1804 born just after the supernova explosion might be a very light NS. However, this may be against the standard supernova explosion theory~\citep{2018MNRAS.481.3305S} \citep[but see][for recent observations]{2022NatAs.tmp..224D}. Thus, our study brings up the issue of how such low-mass NSs are born.
%Thus, RX J1804 born just after the supernova explosion might be a very light NS. \AD{However, this may be against the standard supernova explosion theory~\citep{2018MNRAS.481.3305S} (but see \cite{2022NatAs.tmp..224D} for recent observations). Thus, the mass evolution of RX J1804 }

%, if the scenario that it becomes fat is correct. Hence, RX J1804 could be a site to probe the gap of minimum-mass NSs between supernova explosion theory~\citep{2018MNRAS.481.3305S} and observations~\citep{2022NatAs.tmp..224D} from the mass accretion history.
%\ADcomment{Is it too saying ? If most authors feel so, I delete the last sentence. ($\rightarrow$ All)}

Regarding the EOS dependence, the small radius models tend to have lower $\dot{M}_{{-9},{\rm 2nd}}/\dot{M}_{{-9},{\rm 1st}}-1$. In case of $M_{\rm NS}=1.4~M_{\odot}$, the only Togashi EOS is inconsistent with RX J1804 observations, regardless of the metalicity. This implies a larger radius with $1.4~M_{\odot}$ NSs than 12.7 km, corresponding to that with the LS220 EOS. A similar constraint can be obtained for $1.1~M_{\odot}$ NSs, but it becomes laxer compared with $1.4~M_{\odot}$ NSs. Thus, we suggest that the observational accretion-rate ratio is a new powerful tool to constrain the NS structure.

%%%%%%%%%%%%%%%%%%%%%%%

\section{Conclusions}\label{sec:con}\label{sec:conclusion}

%やったこと
We performed numerical calculations to model a newly-discovered clocked burster RX J1804 for the first time. From the observed recurrence time and persistent flux in two epochs, we found that RX J1804 could become a powerful site to constrain EOSs, even without light-curve modeling. Specifically, small-radius EOSs such as the Togashi are disfavored. This trend is against that of the clocked burster GS 1826$-$24 where large-radius EOSs ($R_{\rm NS}\gtrsim14~{\rm km}$) are disfavored due to the photospheric radius expansion~\citep{2021ApJ...923...64D} \citep[but see also][]{2020MNRAS.494.4576J}. Thus, the combination of GS 1826$-$24 and RX J1804 gives tight constraints on EOSs, i.e., $R_{\rm NS}\sim13~{\rm km}$ in case of $1.4~M_{\odot}$ NSs.

%Future Work
In our burst models, we considered the \textit{standard} energy sources inside accreting NSs, but additional sources may change our results. The one of sources is a shallow heating process inferred from observations \citep[e.g.,][]{2015ApJ...809L..31D}, though the physical origin is unknown. On RX J1804 bursters, some studies indicate the existence of shallow heating source with $\sim0.9~{\rm MeV}$ per accreted nucleon during outburst state~\citep{2017MNRAS.466.4074P, 2018MNRAS.476.2230P}\footnote{This result is obtained with use of \texttt{NSCOOL} code modified for accreting NSs \citep{2013PhRvL.111x1102P}, which was confirmed to be well reproduced with the \texttt{dSTAR} code~\citep{2015ascl.soft05034B} (Rahul Jain, private communication).}. The increase of deep crustal heating rate by $\sim0.9~{\rm MeV/u}$ could highly decrease $\Delta t$, depending on model parameters \citep{2018ApJ...860..147M}. This may reduce the slope in $\dot{M}_{-9}$--$\Delta t$ plane, i.e., $\eta$, leading to higher accretion-rate ratio. The present constraints on EOSs are therefore \textit{minimal} ones, which must be more rigid in the presence of shallow heating.

Not only the heating but also cooling processes may be open such as the strong neutrino Urca cycle in the inner crust~\citep{2014Natur.505...62S}, and the direct Urca process in the core. A wider investigation of present model parameters regarding the energy sources is therefore needed and left for our future work. Nevertheless, we emphasize that energy sources in the crust must be there in all accreting NSs, and the differences of $\Delta t$ among mass/radius might not be so large. Furthermore, the direct Urca process (or more exotic $\nu$ cooling processes) is unlikely to occur according to our conclusion that stiffer NSs, i.e., lower central density, are favored in RX J1804. In these senses, even if the additional energy sources are considered, our conclusion should qualitatively hold.

In this work, we only compared $\Delta t$ and the accretion-rate ratio, but the shape of light curves is needless to say important to constrain model parameters above all for reaction rates, whose uncertainties are reflected on the tail parts~\citep[e.g.,][]{2019ApJ...872...84M}. The burst duration, the time from peak luminosity to the half of peak one, is found to be about 30--40 s for RX J1804~\citep{2019ApJ...887...30F}, which is similar(or a little shorter) to GS 1826$-$24. This must imply high metalicity according to \cite{2007ApJ...671L.141H,2016ApJ...819...46L}~\footnote{In fact, assuming the count-rate to flux conversion factor is independent of energy, our models also indicate $Z_{\rm CNO}\simeq2\%$ from the observed burst duration.}. Moreover, such a long tail of light curves may suggest that rapid-proton capture ($rp$) process is very active to synthesize very heavy proton-rich nuclei with the mass number $\sim100$~\citep{2001PhRvL..86.3471S}. Thus, since the accretion rate is quite high in particular for the 1st epoch, millihertz quasi-periodic oscillation (QPO), which is triggered by marginal stable nuclear burning~\citep{2007ApJ...665.1311H,2014ApJ...787..101K,2023MNRAS.525.2054L}, could occur in RX J1804 and actually has been detected by XMM-Newton observations~\citep{2021MNRAS.500...34T}. Thus, the burst light curves of RX J1804 have many interesting features and must provide much helpful information on nuclear astrophysics. The future work is expected to analyze the burst profiles of RX J1804 to produce the observational light curves comparable with our burst models. We will present the constraints on various model parameters by fitting light curves of RX J1804 elsewhere.

\begin{acknowledgments}
We thank T. Tamagawa for giving useful comments. A.D. also thanks E. Britt, E. Brown, R. Jain, and H. Schatz for fruitful discussion and warm hospitality during his stay at the FRIB. A.D. and N.N. were financially supported by IReNA (International Research Network for Nuclear Astrophysics). This project was financially supported by JSPS KAKENHI (19H00693, 20H05648, 21H01087, 23K19056), RIKEN Pioneering Project ``Evolution of Matter in the Universe (r-EMU), and the RIKEN Incentive Research Project. Parts of the computations were carried out on computer facilities at CfCA in NAOJ and at YITP, Kyoto University.
\end{acknowledgments}

\bibliography{ref}{}
\bibliographystyle{aasjournal}



\end{document}